# Spectroscopy of the Double Quasars Q1343+266AB: The Relationship between Galaxies and QSO C IV Absorption Lines

Arlin P.S. Crotts[1], Jill Bechtold[2], Yihu Fang[1], Robert C. Duncan[3]

## Abstract

We present spectra of QSOs 1343+266A, B, covering Ly $\beta$ to C IV emission at 2Å resolution, and find further evidence that this system is not gravitationally lensed. We see, however, a large number of Ly $\alpha$ and some C IV absorption features common to both spectra. The frequency of C IV lines in common indicates that these absorbers have a radius close to the sightline separation, 39 $h^{-1}$kpc ($q_o = 0.5$), implying a spatial number density of these absorbers the same as current day $L > 0.3 L_*$ galaxies. Furthermore, the velocity mismatch for C IV systems indicates absorber masses in the range $10^{11}\ M_\odot < M < 10^{12}\ M_\odot$. These results confirm that typical galaxies cause QSO C IV-line absorption.

## 1. Introduction

QSO absorption line (QSOAL) systems provide a powerful, nearly unbiased way to study the distribution of baryonic matter in the early Universe, and metal-line absorption systems, containing elements presumably made by stars, offer a method to probe galaxies over all redshifts $0 < z_{abs} \lesssim 5$. Nearly all such information has been obtained by studies of absorbing material along isolated QSO sightlines (e.g. Tytler *et al.* 1987, Sargent *et al.* 1988). With no unique way to disentangle peculiar Doppler shifts from the Hubble flow, these data contain uncertain information about the small-scale spatial clustering and physical size of the absorbers. A primary observational constraint from isolated sightlines is the number of absorbers per unit redshift $N(z)$, but our ignorance of the absorber cross-section stymies attempts to infer the spatial number density of absorbers, $n(z)$.

To introduce size information into QSOAL studies, we can correlate data from adjacent sightlines and thereby introduce sensitivity to the transverse separation between QSOs. The accessible scales are set by the sightline separation, which for proper distance between unlensed QSOs is a nearly constant function of redshift $z \gtrsim 0.5$. The proper sightline separations available for unlensed QSO pairs and groups have tended to hundreds of kpc in proper coordinates (Shaver & Robertson 1983, Crotts 1989) or larger (Jakobsen *et al.* 1986), whereas lensed QSOs (Young *et al.* 1981, Crotts 1988, Magain *et al.* 1988, Steidel & Sargent 1991, Smette *et al.* 1992) probe structure on the scale of only a few kpc. To sense the effects of galaxy size, we must study scales intermediate to these. This supplies not only information about the absorber size and therefore $n(z)$, but also probes individual galaxies at various locations in their rotation curves, leading to an estimate of their mass. It also provides a separate test of the hypothesis that metal absorbers are due to galaxies.

---

[1] Columbia University, Dept. of Astronomy, 538 W. 120th St., New York, NY 10027

[2] Steward Observatory, University of Arizona, Tucson, AZ 85721

[3] University of Texas, Dept. of Astronomy, Austin, TX 78712-1083

Here we discuss our spectroscopy of the one known QSO pair capable of probing galactic scales, the 9.5 arcsec separation binary QSO 1343+266 (Crampton *et al.* 1988). Here we discuss the metal absorption line systems; in a companion paper the Ly $\alpha$ forest results are treated (Bechtold *et al.* 1994, hereafter Paper I).

## 2. Observations and Analysis

We observed Q1343+266A, B on UT 1994 April 4 and 5 with the MMT Blue Spectrograph and 800 l/mm grating. Details of the observing proceedure and reductions are given in Paper I. Figure 1 shows the spectra. Two of us (JB and AC) independently fit splines to the continua and searched for absorption lines, given here in Table 1, and shown in figure 1 of Paper I. The continuum is problematic in the noisy region blueward of 3400Å, and the two linelists differed slightly. These differences were used to estimate the systematic error in wavelength centroid and equivalent width, and this error was added in quadrature with shot-noise. Lines stronger than $3.5\sigma$ were retained, from lists originally constructed with a threshold of $3\sigma$. After examining standard star and sky spectra, we eliminated several weak features likely due to telluric absorption or poorly subtracted sky.

We identified metal-line systems using the comparison list from Morton *et al.* (1988). In addition to identifying lines redward of Ly $\alpha$ emission, we searched for metal lines at every redshift produced by assuming lines blueward of Ly $\alpha$ emission were due to Ly$\alpha$ absorption. We identify the following systems in Q1343+266A: 1) $z_{sys} = 0.5160$ (*definite*): The strongest Mg I and II lines and a strong line of Fe II make this system unmistakable. Additional Fe II lines are expected, but the $S/N$ level of the line seen is not high. The Mg II lines agree within 7 km s$^{-1}$ (versus the expected $1\sigma$ spread of 22 km s$^{-1}$) whereas the Fe II line lands 56 km s$^{-1}$ redward and Mg I 61 km s$^{-1}$ to the blue; 2) $z_{sys} = 1.4003$ (*probable*): Consisting of only a C IV pair, the two lines agree within 13 km s$^{-1}$ and have a doublet ratio of $1.85 \pm 0.21$, consistent with two. The probability of such similarity to a C IV doublet occurring by chance is about $10^{-3}$ redward of Ly $\alpha$; 3) $z_{sys} = 1.5092$ (*definite*): This system is marked by the C IV doublet, plus lines of C II, Si II, Al II, and marginal detections of Fe II $\lambda 1608$ and Al III $\lambda\lambda 1854, 1862$. All except the Al III lines agree within 24 km s$^{-1}$; 4) $z_{sys} = 1.7144$ (*possible*): This system is marked only by Ly $\alpha$ and the *weaker* line of the C IV, a $4.1\sigma$ detection. The $3\sigma$ upper limit on the doublet ratio is 1.3, implying $b < 6$ km s$^{-1}$, which corresponds to $T < 2.7 \times 10^4$ K; 5) $z_{sys} = 2.0342$ (*definite*): The most securely identified of these metal systems; it shows 9 lines of H I, C II, C IV, N V, Si II, Si III and Si IV, and marginal O I $\lambda 1302$ and C I $\lambda 1277$, a wide range of ionization states. The line centroids span 300 km s$^{-1}$ with no obvious correlation with ionization state. This system lands at the QSO redshift, but neutral atomic lines might imply separation from the AGN itself.

In Q1343+266B we find: 1) $z_{sys} = 1.3997$ (*probable*): This system consists of a C IV doublet, with both members at $z = 1.3997$, with a doublet ratio less than two. This may be due to a few narrow, saturated components. The $z = 1.3996$ Si II$\lambda 1526$ line, in the Lyman $\alpha$ forest, may be too strong. 2) $z_{sys} = 1.4191$ (*probable*): This system consists of a C IV pair with a doublet ratio much less than two and both lines at $z = 1.4191 \pm 0.0002$, The weaker component may be significantly broader than the stronger one. 3) $z_{sys} = 1.5110$ (*definite*): This shows a roughly 2-to-1 doublet ratio C IV pair, plus Si II$\lambda 1526$ and Al II$\lambda 1670$. All lines land within about 70 km s$^{-1}$ of each other.

*Other possible C IV doublets*: We also searched for weak line pairs with the CIV wavelength ratios. In A, we found two redward of Ly$\alpha$, at $z_{sys}$=1.4085 and 1.4202; they have low S/N and poorly constrained doublet ratios which are nonetheless far from predicted values. A candidate in the forest is $z_{sys}$=1.2011, which is a good fit to CIV in wavelength and doublet strength ratio. In B, two lines at 3663.54Å and 3669.25Å are consistent with C IV, but such a match is expected by chance at the level of ≈3%. The redder line matches an A spectrum line within 0.01Å, so these are probably associated. The 3669.24Å line in the A spectrum, detected at 28$\sigma$, cannot be due to C IV$\lambda$1550, since $\lambda$1548 is missing. This implies that the lines in B are probably not due to C IV, either.

*BAL trough in Q1343+266A:* A broad, shallow absorption feature is seen blueward of CIV in the spectrum of Q1343+266 A, in addition to narrow associated CIV absorption at z=2.0342. The absorption trough is about 7300 km sec$^{-1}$ wide, begins about 3875 km sec$^{-1}$ bluewards of the peak of the CIV emission, and depresses the continuum by ∼35 %. This feature is also seen in the spectrum shown in Figure 2 of Crampton et al. 1988. No corresponding absorption is seen at SiIV, Ly$\alpha$ or NV, with a continuum drop less than 5%. This CIV trough is unusual in that most weak CIV BAL profiles are clumpy, whereas this one is relatively smooth (cf Turnshek 1988, Weymann et al 1991, Korista et al. 1993). However the spectrum of Q1343+266A does look like the "transition" object UM660 shown by Turnshek 1988. Scaling the relative column densities of HI, CIV, NV, and SiIV given for the well-studied BAL QSO 0226-102 (Korista et al 1992), or for the composite BALs given by Weymann et al (1991) or Korista et al (1993), one might expect a weak BAL wind to be undetectable in NV or SiIV. However, Ly$\alpha$ absorption should still be seen, and our limits on Ly$\alpha$ are about a factor of ∼3 weaker than expected assuming the N(HI)/N(CIV) seen in 0226-102. Given the likely complexity of the regions giving rise to the absorption in BALs (eg Begelman et al 1991, Hamman et al 1993), and the observed variety in the absorption profiles, the CIV/Ly$\alpha$ ratio seen here is extreme but probably can be produced by standard BAL models.

*Q1343+266: Lens or Physical Pair?*: Crampton *et al.* (1988) conclude that Q1343+266 is not lensed based on emission line strength differences between the two spectra and the absence of any lensing galaxy. Our data also strongly indicate that these are physically separate QSOs. Quasar A has strong NV emission, an associated, narrow CIV absorber, and BAL-like absorption shortward of the CIV emission, while B has narrower Ly $\alpha$ than A, weak NV emission, and no BAL. The quotient of the two spectra, also plotted in Figure 1, shows more irregularity than for any QSO lens (e.g. Magain *et al.* 1988, Surdej *et al.* 1988, Steidel & Sargent 1991), despite the pair being only slightly wider than some confirmed lenses.

## 3. Implications for Metal-line Absorber/Galaxy Relationship

Of the four definite or probable metal systems in A and three in B, the systems near $z_{sys} = 1.40$ and 1.51 match, while those at 0.5160 and 2.0342 in A, and 1.4191 in B, do not. Perhaps the two isolated A systems are special cases: the $z_{sys} = 2.0342$ system occurring at the QSO redshift, and 0.5160 being based on Mg II, not C IV. Still, at $z \approx 1.5$, about 50% of the systems are seen to match, indicating a typical absorber radius comparable to the sightline separation, $s = 39$ $h^{-1}$kpc with $q_0 = 0.5$ (50 $h^{-1}$kpc for $q_0 = 0.1$). At larger separations, $s \approx 500$ $h^{-1}$ kpc, no matches are seen (Crotts 1989). Five C IV systems between the Lyman $\alpha$ and C IV emission lines correspond to 3.7 per unit $z$, with a mean $\langle z_{sys} \rangle = 1.70$ and a cutoff $W_0 \approx 0.15$Å. This approximates the Sargent, Boksenberg & Steidel (1988) sample S1, albeit at a lower $\langle z_{sys}\rangle$, 1.70 rather than 2.037, with line of sight density $N = 3.1$ per unit redshift, evolved to $\langle z_{sys}\rangle = 1.70$. These $N$ values are consistent.

Given a rough radius and $N$, we can compute the comoving spatial number density, $n(z)$, for the S1 sample. For $q_0 = 0.5$ (or $q_0 = 0.1$), $N(z = 1.7) = 3.1$ and radius $r_0 = 39\ h^{-1}$kpc (or 50 $h^{-1}$kpc) corresponds to $n = 0.049\ h^3$ Mpc$^{-3}$ (or 0.042 $h^3$ Mpc$^{-3}$). This is also the number density of present-day galaxies brighter than $\approx 0.3\ L_*$ (c.f. Schechter 1976).

This is one of the first indications of the size of metal absorbers sampled by C IV, aside from some information from lensed QSOs still bearing the ambiguity of uncertainties in $z_d$ (Steidel & Sargent 1991). Most data on the size of metal absorbers involve Mg II and Fe II lines, at $z \lesssim 0.5$ in the spectra of QSOs adjacent to absorbing galaxies (Bergeron 1988, Lanzetta & Bowen 1990, Bechtold & Ellingson 1992), and these data indicate galaxy radii of about 50 $h^{-1}$kpc (Lanzetta & Bowen 1990). Our result concerns only C IV absorbers twice as early in the Universe as the Mg II/Fe II sample (for $q_o \approx 0.5$). Nevertheless, the sizes indicated are similar. In fact, integrating over the Schechter luminosity function for $L > 0.3 L_*$ and assuming a constant average surface brightness within a Holmberg radius regardless of luminosity, one calculates an implied ratio of absorber radius to the Holmberg radius of 3.9 (5.1 for $q_o = 0.1$), in close agreement with the value needed for all galaxies in order to explain the number of C IV absorbers (Bergeron 1988).

The absorber mass is estimated using the velocity shift in the two matching absorber pairs. Both $z_{sys} \approx 1.400$ systems are defined primarily by a C IV doublet with both lines giving exactly the same redshift. Accounting for shot noise, the systems are separated by $81 \pm 29$ km s$^{-1}$. For each of the $z_{sys} \approx 1.510$ system, the strongest lines all agree to within about 35 km s$^{-1}$. Incorporating the spread in line centroids and shot-noise errors, the two systems differ by $221 \pm 40$ km s$^{-1}$. The sightline separation implies characteristic masses for the two objects of $1.1 \times 10^{11} M_\odot$ and $8.8 \times 10^{11} M_\odot$, allowing for typical inclination effects ($\langle sin^2\ i\rangle = 0.5$), and assuming that the sightlines intersect opposite halves of each rotation curve. Thus, these objects appear to have the size and comoving number density of typical present-day galaxies, and, established independently, galaxy-sized masses as well.

One observation that might reveal the nature of the two galaxies for which we have estimated the mass (and provide better estimates of the assumed geometric parameters) would be direct imaging of the galaxies in a narrow band corresponding to line emission at each galaxy's redshift, in this case perhaps best done for [O II]$\lambda 3727$ (e.g. Yanny & York 1992). This is an ambitious, but potentially useful, extension of the current results.

In detail we must consider a distribution of masses and sizes, with those subtending sufficient angle to bridge the sightlines tending to be larger than those in general. Thus the size and mass measurements from these data do not sample all parts of the absorber population with the same statistical weight. Nonetheless, size, $n$ and mass are consistent with the population of galaxies larger than a few tenths of $L_*$. We see this as new, confirmatory evidence for the assertion that metal-line systems are caused by galaxies.

This work was supported by NSF grant AST 90-58510 and a gift from Sun microsystems (to JB), AST 91-16390 and the David and Lucile Packard Foundation (to AC), and NSF AST 90-20757 (for RD). We thank the MMTO for allocating time for these observations.

**Figure 1.** Spectra of Q1343+266 A,B and their quotient. The top panel shows the spectrum of A, in raw counts (the response across this wavelength region is fairly flat) as a function of vacuum, geocentric wavelength. The connected tick marks indicate the central wavelengths of lines composing the metal-line systems. The middle panel shows the same information for B. The bottom panel shows the ratio of counts in the two spectra, up to an undetermined multiplicative constant. The peak of the Ly$\alpha$ emission line for B, which extends to 460 counts, has been truncated.

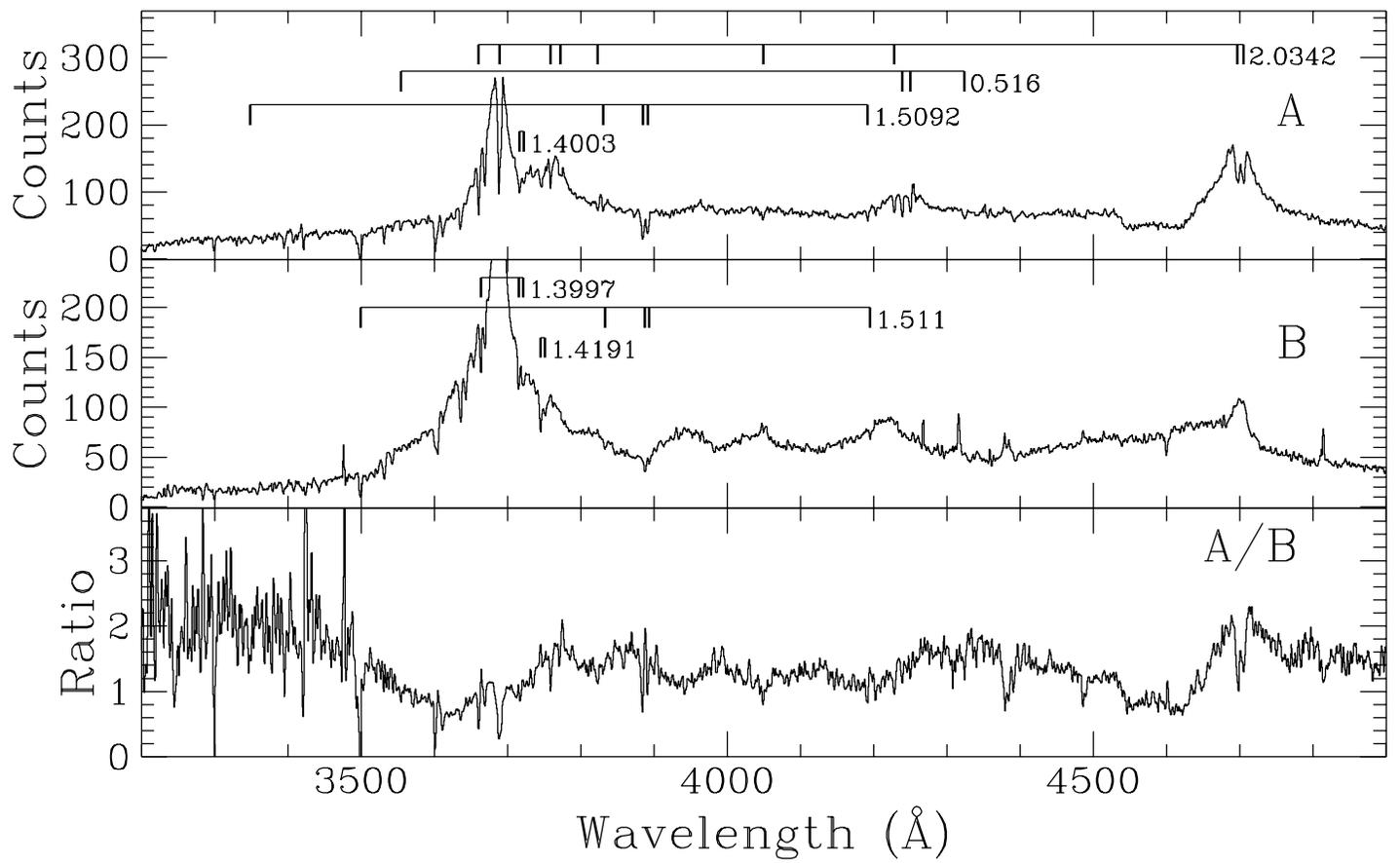

$W$ upper limits are 3.5 sigma confidence.

[1] Ly $\alpha$ component of metal-line absorption system.
[2] Si IV$\lambda$1402 at $z \approx 1.4191$ contributes $W < 0.5$Å.
[3] possible member of C IV doublet, but no non-C IV lines.
[4] Si II$\lambda$1260 at $z \approx 1.7144$ contributes $W < 0.8$Å.
[5] Si IV$\lambda$1393 at $z \approx 1.5090$ contributes $W < 1.2$Å.
[6] Si IV$\lambda$1402 at $z \approx 1.5110$ contributes $W < 0.4$Å.
[7] Fe II$\lambda$2374 at $z \approx 0.5160$ contributes $W < 0.2$Å.
[8] Si II$\lambda$1190 at $z \approx 2.0342$ contributes $W < 0.1$Å, and Fe II$\lambda$2382 at $z \approx 0.5160$ contributes $W < 0.2$Å.
[9] probably not a C IV doublet member, since only one member of this potential doublet has a close match with a non-C IV line in the A spectrum.
[10] Near sky line, so significance and equivalent width may be affected.
[11] C IV$\lambda$1548 ruled out if doublet ratio$> 1.3$.
[12] Si IV$\lambda$1393 to strong with respect to Si IV$\lambda$1402.

TABLE 1
Absorption Lines Detected in QSOs 1343+266A, B

| A: # | $\lambda_c$(Å) | $\sigma_{\lambda_c}$ | $W$(Å) | $\sigma_W$ | S/N | Transition | $z$ | B: # | $\lambda_c$(Å) | $\sigma_{\lambda_c}$ | $W$(Å) | $\sigma_W$ | S/N | Transition | $z$ | A vs. B: $\Delta v$(km/s) |
|---|---|---|---|---|---|---|---|---|---|---|---|---|---|---|---|---|
| 1 | 3218.24 | 0.28 | 1.48 | 0.32 | 4.6 | Ly $\alpha$? | 1.6473 | | | | <1.45 | | | | | |
| | | | <1.05 | | | | | 1 | 3284.15 | 0.25 | 1.72 | 0.36 | 4.8 | Ly $\alpha$? | 1.7015 | |
| 2 | 3299.36 | 0.13 | 1.70 | 0.20 | 8.4 | Ly $\alpha$[1] | 1.7140 | 2 | 3299.65 | 0.26 | 1.55 | 0.39 | 3.9 | Ly $\alpha$? | 1.7143 | $-26 \pm 26$ |
| 3 | 3330.08 | 0.19 | 0.85 | 0.19 | 4.4 | Ly $\alpha$? | 1.7393 | | | | <1.25 | | | | | |
| 4 | 3348.37 | 0.57 | 1.62 | 0.35 | 4.6 | C II 1334 | 1.5090 | | | | <1.39 | | | | | |
| 5 | 3366.39 | 0.24 | 0.86 | 0.21 | 4.1 | Ly $\alpha$? | 1.7692 | | | | <1.09 | | | | | |
| 6 | 3395.12 | 0.13 | 1.75 | 0.18 | 9.6 | Ly $\alpha$? | 1.7928 | 3 | 3393.90 | 0.44 | 1.29 | 0.33 | 3.8 | Ly $\alpha$?[2] | 1.7980 | $107 \pm 40$ |
| 7 | 3407.95 | 0.19 | 1.01 | 0.16 | 6.2 | Ly $\alpha$?[3] | 1.8034 | | | | <1.33 | | | | | |
| 8 | 3413.20 | 0.14 | 0.51 | 0.12 | 4.4 | Ly $\alpha$?[3] | 1.8077 | | | | <1.09 | | | | | |
| 9 | 3421.79 | 0.10 | 1.52 | 0.16 | 9.3 | Ly $\alpha$?[4] | 1.8147 | 4 | 3422.25 | 0.36 | 2.75 | 0.39 | 7.0 | Ly $\alpha$? | 1.8151 | $-40 \pm 33$ |
| 10 | 3444.15 | 0.52 | 1.29 | 0.28 | 4.7 | Ly $\alpha$? | 1.8331 | 5 | 3442.75 | 0.50 | 1.48 | 0.41 | 3.6 | Ly $\alpha$? | 1.8320 | $122 \pm 63$ |
| 11 | 3498.62 | 0.11 | 5.97 | 0.25 | 24.1 | Ly $\alpha$?[5] | 1.8779 | 6 | 3499.17 | 0.14 | 2.63 | 0.23 | 11.3 | Ly $\alpha$?[6] | 1.5106 | $-47 \pm 15$ |
| 12 | 3531.48 | 0.12 | 1.65 | 0.19 | 8.5 | Ly $\alpha$? | 1.9050 | 7 | 3531.48 | 0.21 | 1.41 | 0.20 | 7.1 | Ly $\alpha$? | 1.9050 | $0 \pm 20$ |
| 13 | 3543.06 | 0.17 | 0.52 | 0.11 | 4.9 | Ly $\alpha$? | 1.9145 | 8 | 3542.42 | 0.25 | 0.76 | 0.13 | 5.8 | Ly $\alpha$? | 1.9140 | $54 \pm 25$ |
| 14 | 3554.40 | 0.27 | 0.62 | 0.14 | 4.3 | Fe II 2344 | 0.5162 | | | | <0.37 | | | | | |
| 15 | 3572.02 | 0.20 | 0.41 | 0.10 | 4.2 | Ly $\alpha$? | 1.9383 | | | | <0.37 | | | | | |
| 16 | 3601.71 | 0.09 | 3.64 | 0.16 | 23.4 | Ly $\alpha$?[7] | 1.9627 | 9 | 3603.52 | 0.13 | 2.38 | 0.13 | 18.8 | Ly $\alpha$? | 1.9642 | $-150 \pm 13$ |
| 17 | 3611.93 | 0.13 | 1.82 | 0.14 | 13.2 | Ly $\alpha$?[8] | 1.9711 | 10 | 3612.27 | 0.21 | 0.43 | 0.08 | 5.6 | Ly $\alpha$? | 1.9714 | $-28 \pm 20$ |
| 18 | 3627.07 | 0.35 | 0.57 | 0.12 | 4.7 | Ly $\alpha$? | 1.9836 | | | | <0.21 | | | | | |
| 19 | 3636.05 | 0.11 | 1.55 | 0.11 | 14.5 | Ly $\alpha$? | 1.9910 | 11 | 3635.83 | 0.09 | 1.29 | 0.07 | 18.4 | Ly $\alpha$? | 1.9908 | $18 \pm 12$ |
| | | | <0.36 | | | | | 12 | 3643.15 | 0.12 | 0.74 | 0.06 | 11.9 | Ly $\alpha$? | 1.9968 | |
| 20 | 3653.38 | 0.27 | 0.30 | 0.06 | 4.7 | Ly $\alpha$? | 2.0005 | 13 | 3654.38 | 0.20 | 0.79 | 0.06 | 12.2 | Ly $\alpha$? | 2.0061 | $-82 \pm 27$ |
| 21 | 3660.85 | 0.04 | 1.93 | 0.06 | 34.7 | Si III 1206 | 2.0343 | | | | <0.13 | | | | | |
| | | | <0.17 | | | | | 14 | 3663.54 | 0.06 | 0.80 | 0.04 | 19.6 | Si II 1526[9] | 1.3996 | |
| 22 | 3669.24 | 0.04 | 1.13 | 0.04 | 28.4 | Ly $\alpha$? | 2.0183 | 15 | 3669.25 | 0.09 | 0.51 | 0.04 | 13.1 | Ly $\alpha$?[9] | 2.0183 | $-1 \pm 8$ |
| 23 | 3688.94 | 0.03 | 3.36 | 0.04 | 83.5 | Ly $\alpha$[1] | 2.0345 | | | | <0.05 | | | | | |
| 24 | 3716.19 | 0.15 | 1.72 | 0.10 | 18.1 | C IV 1548 | 1.4003 | 16 | 3715.23 | 0.10 | 0.64 | 0.05 | 12.0 | C IV 1548 | 1.3997 | $77 \pm 14$ |
| 25 | 3722.45 | 0.21 | 0.93 | 0.09 | 10.4 | C IV 1550 | 1.4004 | 17 | 3721.40 | 0.22 | 0.47 | 0.06 | 7.4 | C IV 1550 | 1.3997 | $84 \pm 24$ |
| 26 | 3728.51 | 0.29 | 0.19 | 0.05 | 3.6 | [3] | | | | | <0.15 | | | | | |
| 27 | 3735.62 | 0.53 | 0.32 | 0.08 | 4.2 | [3] | | | | | <0.16 | | | | | |
| 28 | 3746.69 | 0.20 | 1.28 | 0.09 | 15.0 | [3] | | 18 | 3745.39 | 0.12 | 1.28 | 0.09 | 13.5 | C IV 1548 | 1.4192 | $104 \pm 18$ |
| | | | <0.18 | | | | | 19 | 3751.20 | 0.35 | 0.86 | 0.11 | 7.6 | C IV 1550 | 1.4189 | |
| 29 | 3753.79 | 0.13 | 0.19 | 0.04 | 4.9 | [3] | | | | | <0.18 | | | | | |
| 30 | 3758.97 | 0.08 | 0.91 | 0.05 | 16.8 | N V 1238 | 2.0343 | | | | <0.23 | | | | | |
| 31 | 3772.28 | 0.30 | 0.62 | 0.08 | 7.9 | N V 1242 | 2.0353 | | | | <0.25 | | | | | |
| 32 | 3823.27 | 0.31 | 0.39 | 0.09 | 4.3 | Si II 1260 | 2.0333 | | | | <0.25 | | | | | |
| 33 | 3830.83 | 0.12 | 0.26 | 0.06 | 4.5 | Si II 1526 | 1.5092 | 20 | 3832.98 | 0.37 | 0.37 | 0.10 | 3.6 | Si II 1526 | 1.5106 | $-168 \pm 30$ |
| 34 | 3884.77 | 0.11 | 2.62 | 0.13 | 19.8 | C IV 1548 | 1.5092 | | | | <0.43 | | | | | |
| | | | <0.32 | | | | | 21 | 3887.67 | 0.22 | 1.29 | 0.16 | 8.3 | C IV 1548 | 1.5111 | |
| 35 | 3891.40 | 0.11 | 1.75 | 0.12 | 15.0 | C IV 1550 | 1.5093 | | | | <0.38 | | | | | |
| | | | <0.32 | | | | | 22 | 3893.77 | 0.31 | 0.59 | 0.13 | 4.5 | C IV 1550 | 1.5109 | |
| | | | <0.32 | | | | | 23 | 3982.65 | 0.49 | 0.50 | 0.14 | 3.6 | C IV 1548 | 1.5724 | |
| 36 | 4049.10 | 0.26 | 0.66 | 0.12 | 5.3 | C II 1334[10] | 2.0341 | | | | <0.20 | | | | | |
| 37 | 4192.15 | 0.17 | 0.61 | 0.09 | 6.8 | Al II 1670 | 1.5091 | | | | <0.26 | | | | | |
| | | | <0.24 | | | | | 24 | 4195.61 | 0.26 | 0.34 | 0.08 | 4.4 | Al II 1670 | 1.5112 | |
| 38 | 4203.05 | 0.27 | 0.38 | 0.09 | 4.1 | C IV 1550?[11] | 1.7148 | | | | <0.20 | | | | | |
| 39 | 4228.54 | 0.12 | 0.78 | 0.07 | 10.9 | Si IV 1393[12] | 2.0339 | | | | <0.18 | | | | | |
| 40 | 4239.25 | 0.15 | 0.99 | 0.11 | 8.7 | Mg II 2796 | 0.5160 | | | | <0.39 | | | | | |
| 41 | 4250.04 | 0.31 | 0.68 | 0.15 | 4.6 | Mg II 2803 | 0.5160 | | | | <0.29 | | | | | |
| 42 | 4324.10 | 0.44 | 0.80 | 0.15 | 5.5 | Mg I 2852[10] | 0.5160 | | | | <0.25 | | | | | |
| | | | <0.27 | | | | | 25 | 4331.00 | 0.67 | 0.72 | 0.18 | 4.0 | [10] | | |
| 43 | 4588.71 | 0.17 | 0.34 | 0.09 | 3.6 | [10] | | | | | <0.51 | | | | | |
| 44 | 4681.33 | 0.19 | 0.19 | 0.04 | 4.3 | [10] | | | | | <0.59 | | | | | |
| 45 | 4697.39 | 0.15 | 2.09 | 0.10 | 21.4 | C IV 1548 | 2.0341 | | | | <0.50 | | | | | |
| 46 | 4705.43 | 0.12 | 1.50 | 0.08 | 18.9 | C IV 1550 | 2.0342 | | | | <0.50 | | | | | |
| 47 | 4751.39 | 0.22 | 0.34 | 0.08 | 4.1 | [10] | | | | | <0.37 | | | | | |